\documentclass[12pt]{article}

\usepackage{eurosym}
\usepackage{graphicx}
\usepackage[utf8]{inputenc}
\usepackage[T1]{fontenc}
\usepackage{indentfirst}
\usepackage[margin=0.6in,nomarginpar]{geometry}
\usepackage[final]{hyperref}
\usepackage{amsmath}
\usepackage{hyperref}
\usepackage{cite}
\usepackage{subcaption}
\usepackage{caption}
\usepackage{amssymb}
\usepackage{multirow}
\usepackage[table]{xcolor}
\usepackage{orcidlink}
\hypersetup{
colorlinks=true,
linkcolor=blue,
citecolor=blue,
filecolor=magenta,
urlcolor=blue
}

\begin{document}

\title{The ideal gas of Bosons and Fermions in Harmonic Traps 

in the framework of Extended Uncertainty Principle}
\author{B. Hamil \orcidlink{0000-0002-7043-6104} \thanks{hamilbilel@gmail.com} \\
Laboratoire de Physique Math\'{e}matique et Subatomique,\\
Facult\'{e} des Sciences Exactes, Universit\'{e} Constantine 1 Fr\`{e}res Mentouri, \\Constantine,
Algeria. \and B. C. L\"{u}tf\"{u}o\u{g}lu \orcidlink{0000-0001-6467-5005}\thanks{%
bekir.lutfuoglu@uhk.cz (corresponding author) } \\
Department of Physics, Faculty of Science,  University of Hradec Kralove,\\
Rokitanskeho 62/26,  Hradec Kralove, 500 03, Czechia.}
\date{\today }
\maketitle

\begin{abstract}
This manuscript studies harmonically trapped ideal Bose and Fermi gas systems and their thermodynamics in the framework of the Extended Uncertainty Principle (EUP). In particular, we demonstrated how the ground and thermal particle ratios, condensate temperature, internal energy, specific heat, and equation of state functions change in the EUP formalism.  Following a comprehensive analysis, we concluded that the effect of the EUP on ideal Bose and Fermi gas systems remains relatively modest compared to the effect of the Gravitational (Generalised) Uncertainty Principle. { By comparing the obtained results with experimental data, we found that the  EUP parameter is bounded as $\alpha \leq 0.36654\times 10^{7}\text{ \ }m^{-2}$.}
\end{abstract}


\section{Introduction}

Although the need for a lower bound on the measurable length of the order of the Planck length scale became popular at the end of the last century, especially in connection with string theory \cite{Veneziano, Witten, Amati} and loop quantum gravity \cite{Maggiore}, its history goes back much further \cite{Snyder}. { During those years, when the debate was at its strongest, Kempf et al. adapted the idea of minimal length to quantum mechanics by modifying the standard Heisenberg algebra and applying particular modifications to the canonical commutation relations between position and momentum operators, thus changing the structure of the Heisenberg algebra as follows \cite{Math, Phys, Kempf, Hinrichsen1996}:
\begin{equation}
\left[ X,P\right] =i\hbar \left( 1+\beta P^{2}\right),
\end{equation}%
This modification has been called the generalized uncertainty principle (GUP). 
}

{On the other hand, quantum field theory on curved spacetime represents a fundamental challenge at the intersection of general relativity and quantum theory. In curved spacetime, the concept of a plane wave becomes ill-defined, leading to limitations in the precision of momentum measurements, which can be expressed as a non-zero minimal uncertainty in momentum. This suggests that traditional quantum mechanical frameworks may require modifications when applied to curved geometries. In addition, in \cite{salvator}, Mignemi showed that on (anti-)de Sitter backgrounds, the Heisenberg uncertainty principle must be modified by introducing a small correction. This modification is known as the extended uncertainty principle (EUP) \cite{Bolen, Park}. The extended commutation relations introduced by Mignemi on anti-de Sitter backgrounds describe this modification:
\begin{equation}
\left[ X,P\right] =i\hbar \left( 1+\alpha X^{2}\right),
\end{equation}%
where $\alpha=\frac{\alpha_{0}}{L}$, and $\alpha_{0}$ is a dimensionless constant of order unitary, while $L$ is a constant parameter \cite{Mureika}.
}
After these important studies, quantum deformations and their consequences were studied in many studies within various fields of physics \cite{Nozari2006, Fityo, Vakili2012, Matin, Zhang2015, Filho, Lambiase2017,  Hamil2019, Luciano2019, Dabrowski2020, BB2021, Hamil2021, BBM2022, Li2022, Hao2022, Boudjema, BBA2023, Oubagha2023, Lambiase2023, Senay2023, Benhadjira2024, Wojnar2024, Nozari2024}.

In the fields of statistical physics and thermodynamics, the influence of gravity is typically disregarded. However, the generalized and extended uncertainty principles, derived from the quantum theory of gravitation, necessitate a modification to the density of quantum states in statistical physics, which may occasionally result in a non-negligible impact on the properties of matter \cite{Fityo, Vakili2012}. This consequently results in a change to the thermal properties of thermodynamic systems. In this framework, Zhang et al. discussed how the minimal length effect can modify the condensation in a relativistic ideal Bose gas system in \cite{Zhang2015}. Later, Li et al. and Senay analyzed Bose and Fermi gas systems under the GUP and provided the resulting corrections for thermal quantities in references \cite{Li2022} and \cite{Senay2023},  respectively. Wojnar  and Benhadjira et al. have recently investigated the GUP and EUP effects on Bose-Einstein condensates (BEC) \cite{Wojnar2024, Benhadjira2024}. It is important to emphasize that these modifications can appear not only in the high-temperature regime but also in the low-temperature regime as discussed in \cite{Boudjema, Lili}. 

The application of an external potential represents a significant methodology for the confinement and probing of quantum gases. Modifying the intensity and shape of the external potential enables the control of diverse behaviors exhibited by the quantum system, thereby facilitating the occurrence of a phase transition at low temperatures. Consequently, potential constraints exert a profound influence on the properties of quantum systems. Therefore, the thermodynamic properties of quantum gases have been investigated in numerous research papers \cite{Su2006, Grether2007, Du2012}.  
Further, the examinations have been extended by considering the Planck scale effects \cite{BrisceseEPL2012, Brisceseetal2012, Castellanos2015,  HLi2018, Ya}.

In an interesting work, Klukikov et al. studied an ideal quantum Fermi gas in curved spacetime and showed the curvature affects to the chemical potential and Fermi energy \cite{Kulikov1995}. All these facts prompted us to investigate the curvature effects on the thermal properties of ideal boson and fermion gases constrained by the isotropic harmonic potential within the EUP framework. 
To achieve these goals, we  construct the manuscript as follows: in sec. \ref{sec2} we handle an ideal Bose gas system and derive its EUP-modified thermo-statistical properties, including the ground, excited particle number ratios, condensate temperature ratio, internal energy and specific heat functions, and equation of state functions. Then, in sec. \ref{sec3} we take into account an ideal Fermi gas and examine its EUP-modified thermo-statistical properties in terms of the similar quantities. Moreover, in sec. \ref{subsec1}, we delve in the degenerated case. 
We conclude the manuscript in sec. \ref{conc}.

\section{Modified Ideal Bose-Einstein gas  in Harmonic Traps}\label{sec2}

Let us examine a Bose gas consisting of $N$ neutral atoms that are trapped within a three-dimensional isotropic harmonic oscillator potential described by%
\begin{equation}
V\left( r\right) =\frac{m\omega ^{2}}{2}r^{2},
\end{equation}%
where $m$ denotes the mass of the atoms, while $\omega$ corresponds to their trap frequencies. In this case, the total energy of the entire system can be calculated by adding up the individual energies of each particle,%
\begin{equation}
E_{n_{1},n_{2},n_{3}}=\hslash \omega \left( n_{1}+n_{2}+n_{3}\right) +E_{0},
\end{equation}
for $n_{i}=0,1,2,...$ and with the zero-point energy,  $E_{0}=\frac{3\hslash \omega}{2}$. In grand canonical statistics, the total number of the system is given by
\begin{equation}
N=\sum_{n_{1},n_{2},n_{3}}\frac{1}{z^{-1}e^{\frac{\hbar \omega \left(
n_{1}+n_{2}+n_{3}\right) }{K_{B}T}}-1}.  \label{tot}
\end{equation}
Here, $K_{B}$ denotes the Boltzmann constant, $T$ is the temperature of the system, $z=e^{\frac{\mu -E_{0}}{K_{B}T}}$ is the fugacity where $\mu $ represents the chemical potential. When the number of trapped particles is
sufficiently large, and the average kinetic energy of a boson particle is
significantly higher than the spacing between energy levels, the energy of
the system can be treated as quasi-continuous. In this scenario, we replace
the summation with ordinary integrals weighted by an appropriate density of
states, denoted as $\rho \left( E\right) $. By adopting this approach, Eq. (%
\ref{tot}) transforms into the following expression%
\begin{equation}
N=N_{0}+N_{e},  \label{te2}
\end{equation}%
where%
\begin{equation}
N_{0}=\frac{z}{1-z},
\end{equation}%
is the number of particles in the ground state and 
\begin{equation}
N_{e}=\int \frac{\rho \left( E\right) dE}{z^{-1}e^{\frac{E}{K_{B}T}}-1}. \label{bcl1}
\end{equation}%
is the number of particles in the thermal states.
{To the best of our knowledge, within a deformed formalism, there are two
methods for deriving the density of states  and the total number of
particles:
\begin{enumerate}

\item  Utilize deformed commutation relations incorporating the EUP along the standard Hamiltonian. In this approach, it is necessary to use deformed measures of integrals, leading to a modification of the density of states \cite{Fityo, PITAEVSKII}.

\item Apply standard commutation relations while using a modified Hamiltonian \cite{Bosso}.

\end{enumerate}
In this work, we opt for the first method. Thus, in the presence of the EUP, the density of states is,
}
\begin{equation}
\rho \left( E\right) =\frac{1}{h^{3}}\int \frac{d\overrightarrow{p}\text{ }d%
\overrightarrow{r}}{\left( 1+\alpha r^{2}\right) ^{3}}\text{ }\delta \left(
E-\frac{p^{2}}{2m}-\frac{m\omega ^{2}}{2}r^{2}\right) .
\end{equation}%
After the straightforward algebra, we derive the density of states
function as%
\begin{equation}
\rho \left(E \right) =\frac{1}{2}\frac{E^{2}}{\hbar ^{3}\omega
^{3}}\left( 1-\frac{3\alpha E}{m\omega ^{2}}\right) .  \label{dens}
\end{equation}%
By utilizing Eq. (\ref{dens}), we transform Eq. (\ref{tot}) 
 into%
\begin{equation}
N=N_0+\frac{1}{2}\left( \frac{K_{B}T}{\hbar \omega }\right) ^{3}\Bigg[
\int_{0}^{+\infty }\frac{y^{2}}{z^{-1}e^{y}-1}dy-\frac{3\alpha K_{B}T}{%
m\omega ^{2}}\int_{0}^{+\infty }\frac{y^{3}}{z^{-1}e^{y}-1}dy\Bigg].\label{8}
\end{equation}%
where $y=\frac{E}{K_{B}T}.$ One may see in Eq. (\ref{8}) the appearance of Bose
function $g_{n}\left( z\right) $%
\begin{equation}
g_{n}\left( z\right) =\frac{1}{\Gamma \left( n\right) }\int_{0}^{+\infty }%
\frac{y^{n-1}}{z^{-1}e^{y}-1}dy,
\end{equation}%
where $\Gamma \left( n\right) $ being the Euler Gamma function. After
integration, we have%
\begin{equation}
N_e=\left( \frac{K_{B}T}{\hbar \omega }\right) ^{3}g_{3}\left(
z\right) \left( 1-\frac{9\alpha K_{B}T}{m\omega ^{2}}\frac{g_{4}\left(
z\right) }{g_{3}\left( z\right) }\right),  \label{nnn0}
\end{equation}
which sets a condition for temperature
\begin{eqnarray}
T \leq \frac{m\omega ^{2}}{9\alpha K_{B}}\frac{g_{3}\left( z\right) }{g_{4}\left(z\right) }.
\end{eqnarray}
Consequently, the total number of particles reads
\begin{equation}
N=\frac{z}{1-z}+\left( \frac{K_{B}T}{\hbar \omega }\right) ^{3}g_{3}\left(
z\right) \left( 1-\frac{9\alpha K_{B}T}{m\omega ^{2}}\frac{g_{4}\left(
z\right) }{g_{3}\left( z\right) }\right) .  \label{nnn}
\end{equation}
It is worth noting that in the limit $\alpha \rightarrow 0$ the EUP correction term drops out and we get the standard total number of particles of the ideal Bose gas.  In addition, in this limit, the condition that determines the upper value of the temperature disappears.

We now aim to explore the phenomenon of Bose condensation by setting the independent variables $N$ and $T$. Here, we consider two conditions: 
\begin{itemize}
    \item If $N \leq N_{e}^{\max }$, then all particles can occupy excited states. Since $z<1$ and we  need to determine  it from the equation 
\begin{equation}
N=\left( \frac{K_{B}T}{\hbar \omega }\right) ^{3}g_{3}\left( z\right) \left(
1-\frac{9\alpha K_{B}T}{m\omega ^{2}}\frac{g_{4}\left( z\right) }{%
g_{3}\left( z\right) }\right) .  \label{NN}
\end{equation}%
In this scenario, we can neglect $N_{0}.$ 

\item If $N>N_{e}^{\max }$, then
the excited states are insufficient to accommodate all the particles%
\begin{equation}
N_{0}=N-\left( \frac{K_{B}T}{\hbar \omega }\right) ^{3}\zeta \left( 3\right)\left( 1-\frac{9\alpha K_{B}T}{m\omega ^{2}}\frac{\zeta \left( 4\right) }{\zeta \left( 3\right) }\right) .\label{bnn}
\end{equation}
To find the EUP-corrected condensation temperature of the ideal Bose gas,  $T_{c}^{EUP}$, we set $N_{0}=0$ and simultaneously let $z=1$. Then, we estimate $T_{c}^{EUP}$ as 
\begin{equation}
\frac{T_{c}^{EUP}}{T_{c}^{B}}= 1+\frac{3\hbar \alpha }{m\omega }\frac{%
\zeta \left( 4\right) }{\zeta \left( 3\right) }\left( \frac{N}{\zeta \left(
3\right) }\right) ^{\frac{1}{^{3}}} ,  \label{tepcc}
\end{equation}%
where 
\begin{eqnarray}
T_{c}^{B}=\frac{\hbar \omega }{K_{B}}\left( \frac{N}{\zeta \left(
3\right) }\right) ^{\frac{1}{^{3}}},     \label{tradcondtemp}
\end{eqnarray}
is the traditional condensation temperature. Eq. (\ref{tepcc}) reveals that the presence of EUP increases the condensation temperature of an ideal Bose gas trapped by a three-dimensional harmonic oscillator. To illustrate this amplification effect, we depict the rate of condensation temperature versus the induced curvature parameter in Figure \ref{fig1}.

\begin{figure}[htb!]
\centering\includegraphics[scale=0.5]{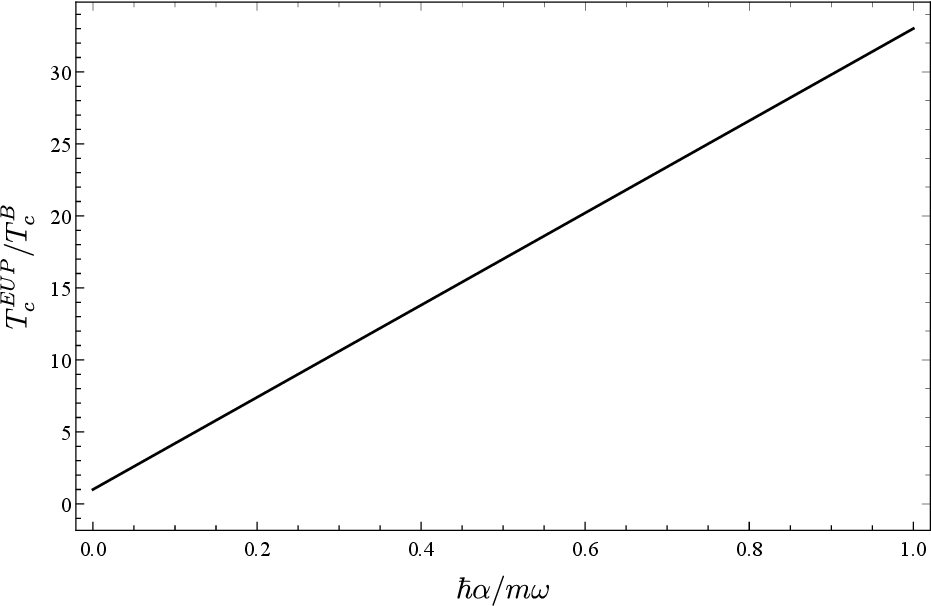}
\caption{The variation of $\frac{T_{c}^{EUP}}{T_{c}^{B}}$ versus $\protect%
\alpha .$}
\label{fig1}
\end{figure}
We observe that the condensation temperature rate increases monotonically as the curvature effect increases. 
   
\end{itemize}

\begin{figure}[htb!]
\begin{minipage}[t]{0.5\textwidth}
        \centering
        \includegraphics[width=\textwidth]{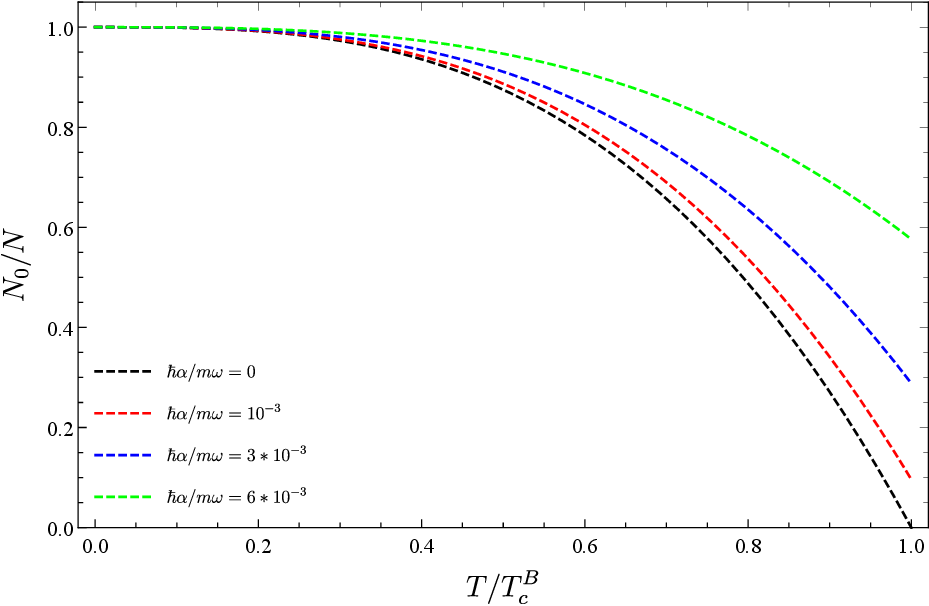}
        \label{fig:ta}
        \subcaption{Ground state}
\end{minipage}%
\begin{minipage}[t]{0.5\textwidth}
        \centering
        \includegraphics[width=\textwidth]{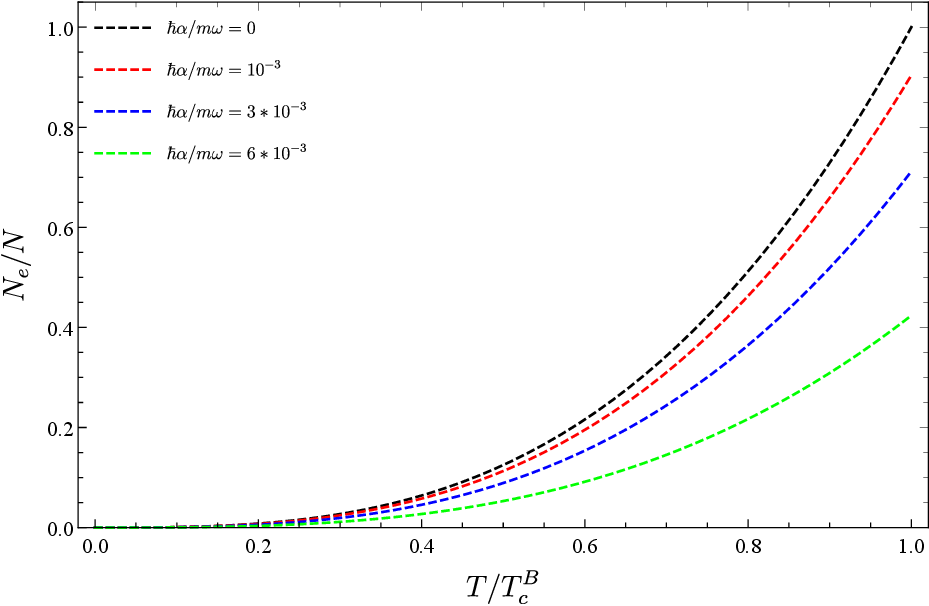}
       \label{fig:tb}
       \subcaption{Thermal state}
   \end{minipage}%
\caption{The effect of the EUP parameter on the variations in the particle number ratio versus $T / T_{c}^{B} $.} 
\label{fig2}
\end{figure}

We now examine the number ratio of particles in the ground state to total particle, $\frac{N_{0}}{N}$. Using Eqs. (\ref{nnn}) and (\ref{bnn}), we  express this ratio as a function of temperature as 
\begin{equation}
\frac{N_{0}}{N}=1-\left(  \frac{T}{T_{c}^{B}}\right) ^{3}\left[1-\frac{9\hbar \alpha }{m\omega }\frac{\zeta \left( 4\right) }{\zeta \left( 3\right) }\left( \frac{N}{\zeta \left( 3\right) }\right) ^{\frac{1}{^{3}}}\right] .
\end{equation}
Here, we note that this ratio is non-zero only when the condition of temperature $T_{c}^{EUP}>T$ is satisfied. We then find the number ratio of the particles in the thermal state to the whole system particles
\begin{equation}
\frac{N_{e}}{N}=\left(  \frac{T}{T_{c}^{B}}\right) ^{3}\left[1-\frac{9\hbar \alpha }{m\omega }\frac{\zeta \left( 4\right) }{\zeta \left( 3\right) }\left( \frac{N}{\zeta \left( 3\right) }\right) ^{\frac{1}{^{3}}}\right] .
\end{equation}
In Fig. \ref{fig2}, we plot these ratios versus temperature ratios for different values of the EUP parameter.

We observe that in both cases the impact of the EUP is negligible for  $T<<T_{c}^B$. When $T<T_{c}^B$, we notice that with higher EUP parameter values the ground state particle number becomes greater while the thermal state particle number becomes smaller.

Next, we study the internal energy of the system. Here, we first consider the relation $U=\sum_{i} N_{i} E _{i}$, and then using the assumption given above we convert the sum into the integral form. After taking the integral we get the EUP-corrected internal energy in the following form:
\begin{equation}
U=3\frac{K_{B}^{4}T^{4}}{\hbar ^{3}\omega ^{3}}g_{4}\left( z\right) \left( 1-%
\frac{12\alpha K_{B}T}{m\omega ^{2}}\frac{g_{5}\left( z\right) }{g_{4}\left(
z\right) }\right) .  \label{een}
\end{equation}
Here the EUP correction term appears as the second term in parentheses. We now consider the case $T\leq T_{c}^{EUP}$ where $z=1$ and $N_{0}\neq 0$. In this case, the internal energy is
\begin{equation}
U=3\frac{K_{B}^{4}T^{4}}{\hbar ^{3}\omega ^{3}}\zeta \left( 4\right) \left(
1-\frac{12\alpha K_{B}T}{m\omega ^{2}}\frac{\zeta \left( 5\right) }{\zeta
\left( 4\right) }\right) ,  \label{21}
\end{equation}
and it means that the internal energy of the ideal Bose gas system decreases within the EUP formalism.  In the other case $T\geq T_{c}^{EUP}$, where $z\neq 1$ and $N_{0}=0,$ we can eliminate the term $\left( \frac{K_{B}T}{\hbar\omega }\right) ^{3}$ in Eq. \eqref{een} with the help of Eq. \eqref{NN}. We then obtain the internal energy in the form of
\begin{equation}
U=3NK_{B}T\frac{g_{4}\left( z\right) }{g_{3}\left( z\right) }\left( 1-\frac{%
12\alpha K_{B}T}{m\omega ^{2}}\frac{g_{5}\left( z\right) }{g_{4}\left(
z\right) }+\frac{9\alpha K_{B}T}{m\omega ^{2}}\frac{g_{4}\left( z\right) }{%
g_{3}\left( z\right) }\right) .  \label{22}
\end{equation}
On the other hand, it is a known fact that the classical limit corresponds to low particle densities and high-temperature conditions. In this limit, the fugacity is very small, $z<<1$, and thus, $g_{n}\left( z\right) \simeq z$. Therefore, within this limit  Eq. (\ref{22}) reduces to the following form
\begin{equation}
U=3NK_{B}T\left( 1-\frac{3\alpha K_{B}T}{m\omega ^{2}}\right) .
\end{equation}
Here, the first term, $3NK_{B}T$, represents the standard internal energy, while the second term, which is quadratic in temperature $-\frac{9N\alpha K_{B}^{2}T^{2}}{m\omega ^{2}}$, is the correction due to EUP. Next, we calculate the heat capacity, using its thermodynamic definition
\begin{equation}
C_{V}=\frac{\partial U }{\partial T}\bigg\vert _{N,V}.\label{cvv}
\end{equation}%
For the case $T<{T_{c}^{B}}$, we easily obtain 
\begin{equation}
\frac{C_{V}}{K_{B}}=12\left( \frac{K_{B}T}{\hbar \omega }\right) ^{3}\zeta
\left( 4\right) \left( 1-\frac{15\alpha K_{B}T}{m\omega ^{2}}\frac{\zeta
\left( 5\right) }{\zeta \left( 4\right) }\right) .
\end{equation}%
However, in the case of $T>{T_{c}^{B}}$ we have to take into account the temperature derivative of the fugacity. So we have
\begin{eqnarray}
C_{V} &=&12\frac{K_{B}^{4}T^{3}}{\hbar ^{3}\omega ^{3}}g_{4}\left( z\right)
\left( 1-\frac{15\alpha K_{B}T}{m\omega ^{2}}\frac{g_{5}\left( z\right) }{%
g_{4}\left( z\right) }\right) +  3\frac{K_{B}^{4}T^{4}}{\hbar ^{3}\omega ^{3}}g_{3}\left( z\right) \left( 1-%
\frac{12\alpha K_{B}T}{m\omega ^{2}}\frac{g_{4}\left( z\right) }{g_{3}\left(
z\right) }\right) \frac{1}{z}\frac{\partial z}{\partial T}.  \label{che}
\end{eqnarray}
Here. we obtain the  quantity $\frac{1}{z}\frac{\partial z}{\partial T}$ by recalling that $N_{0}=0$ and the total particle number is a constant. So we use $\frac{d}{dT}N=0$ and get
\begin{equation}
\frac{1}{z}\frac{\partial z}{\partial T}=-\frac{3}{T}\frac{g_{3}\left(
z\right) }{g_{2}\left( z\right) }\left( 1-\frac{12\alpha K_{B}T}{m\omega ^{2}%
}\frac{g_{4}\left( z\right) }{g_{3}\left( z\right) }+\frac{9\alpha K_{B}T}{%
m\omega ^{2}}\frac{g_{3}\left( z\right) }{g_{2}\left( z\right) }\right) .
\label{mm}
\end{equation}%
We then use Eq. \eqref{mm} in Eq. \eqref{che}, and for  $T>T_{c}^{EUP}$  we express the EUP-corrected heat capacity  function in the following definite form 
\begin{eqnarray}
C_{V} &=& 12 K_{B} N \frac{g_{4}\left( z\right) }{g_{3}\left( z\right) }\left( 1-\frac{15\alpha K_{B}T}{m\omega ^{2}}\frac{g_{5}\left( z\right) }{g_{4}\left( z\right) }+\frac{9\alpha K_{B}T}{m\omega ^{2}}\frac{g_{4}\left(z\right) }{g_{3}\left( z\right) }\right)   \notag \\
&-& 9 K_{B} N\frac{g_{3}\left( z\right) }{g_{2}\left( z\right) }\left( 1-\frac{15\alpha K_{B}T}{m\omega ^{2}}\frac{g_{4}\left( z\right) }{g_{3}\left(z\right) }+\frac{9\alpha K_{B}T}{m\omega ^{2}}\frac{g_{3}\left( z\right) }{g_{2}\left( z\right) }\right) . \label{cancv1}
\end{eqnarray}
In classical limit, Eq. \eqref{cancv1} reduces to 
\begin{equation}
C_{V}=3K_{B}N\left( 1-\frac{6\alpha K_{B}T}{m\omega ^{2}}\right) .
\label{hcla}
\end{equation}%
Here, the first term is the standard heat capacity, while the second term, which linearly depends on temperature $-\frac{18\alpha NK_{B}^{2}T}{m\omega ^{2}}$, is the EUP correction.

Now, we establish the equation of the state of the ideal Bose gas trapped by the harmonic potential in the presence
of EUP. As usual, we start from the thermodynamic definition
\begin{equation}
P=\frac{K_{B}T}{V}\ln \mathcal{Z}\left( \beta ,V,z\right) =-\frac{K_{B}T}{V}%
\sum_{n_{1},n_{2},n_{3}}\ln \left( 1-ze^{-\beta
E_{n_{1},n_{2},n_{3}}}\right) .
\end{equation}
where $\beta=\frac{1}{K_{B}T} $. Here we repeat the approach used above:  First we separate the ground state particle numbers from the total number of particles, and then we approximate the number of thermal particle terms by replacing the summation with an integral. We then get
\begin{equation}
P=-\frac{K_{B}T}{V}\ln \left( 1-z\right) -\frac{K_{B}T}{2V\hbar ^{3}\omega^{3}}\int_{0}^{\infty }E^{2}\left( 1-\frac{3\alpha E}{m\omega ^{2}}\right) \ln \left( 1-ze^{-\frac{E}{K_{B}T}}\right) dE.  \label{par}
\end{equation}%
We note that the existing integral in the second term of  Eq. \eqref{par} is a typical integral of physics and can be solved easily. Following  the necessary algebra, we obtain
\begin{equation}
P=\frac{K_{B}T}{V}\ln \left( \frac{1}{1-z}\right) +\frac{K_{B}T}{V}\left(\frac{K_{B}T}{\hbar \omega }\right) ^{3}g_{4}\left( z\right) \left( 1-\frac{9\alpha K_{B}T}{m\omega ^{2}}\frac{g_{5}\left( z\right) }{g_{4}\left(z\right) }\right) . \label{magda1}
\end{equation}%
In the thermodynamic limit, the first term vanishes and Eq. \eqref{magda1} reduces to 
\begin{equation}
\frac{PV}{K_{B}T}=\left( \frac{K_{B}T}{\hbar \omega }\right) ^{3}g_{4}\left(z\right) \left( 1-\frac{9\alpha K_{B}T}{m\omega ^{2}}\frac{g_{5}\left(z\right) }{g_{4}\left( z\right) }\right) .
\end{equation}
For the case $T < T_{c}^{EUP}$, where  $z=1$ and $N_{0}\neq 0,$ the equation of state becomes
\begin{equation}
\frac{PV}{K_{B}T}=\left( \frac{K_{B}T}{\hbar \omega }\right) ^{3}\zeta \left( 4\right) \left( 1-\frac{9\alpha K_{B}T}{m\omega ^{2}}\frac{\zeta \left( 5\right) }{\zeta \left( 4\right) }\right) ,  \label{25}
\end{equation}
and it demonstrates a decrease in pressure as the EUP parameter increases. 

On the other hand, in the case $T > T_{c}^{EUP}$, where $z\neq 1$, and $N_{0}=0$, Eq. \eqref{magda1} reduces to 
\begin{equation}
\frac{PV}{K_{B}T}=N\frac{g_{4}\left( z\right) }{g_{3}\left( z\right) }\left(1-\frac{9\alpha K_{B}T}{m\omega ^{2}}\frac{g_{5}\left( z\right) }{g_{4}\left( z\right) }+\frac{9\alpha K_{B}T}{m\omega ^{2}}\frac{g_{4}\left(z\right) }{g_{3}\left( z\right) }\right), \label{bp}
\end{equation}%
and in the classical limit, it gives the standard equation of state expression
\begin{equation}
PV=NK_{B}T.
\end{equation}
which is not affected by the EUP.  However, it is important to approximate, albeit roughly, a minimum value for the uncertainty in the momentum.  We can achieve this goal by examining how it affects the heat capacity at constant pressure. Here, we use \cite{Matin} as our guide for this analysis. We use Eq. \eqref{hcla} 
\begin{equation}
\frac{C_{V}-C_{V}^{\alpha =0}}{C_{V}^{\alpha =0}}=-\frac{6\alpha K_{B}T}{%
m\omega ^{2}}.
\end{equation}
Current experimental methods can measure specific heat with an accuracy of approximately $10^{-7}$ \cite{Matin, Mohr}. When we consider hydrogen atoms at room temperature (300K), we can use this level of precision along with Eq. \eqref{hcla}, to establish an upper limit for the EUP parameter
\begin{equation}
\alpha \leq 0.36654\times 10^{7}\text{ \ }m^{-2}.\label{alphaval}
\end{equation}
{ This assumption was also considered in determining the upper bound of the generalized uncertainty principle parameter \cite{Miraboutalebi}. It is important to underline that our finding, expressed in equation (\ref{alphaval}), aligns with the results presented in \cite{bilelmerad, Mirza}.} Note that the magnitude of the particle mass, the Boltzmann constant, the critical temperature, and Fermi energy are typically small valued quantities. Therefore, the magnitude of the first-order correction term of the EUP is significantly larger than that of the second-order correction term, leading to the consideration of only the first-order correction term of the EUP. Here we follow \cite{Ya} and use two Bose gases as examples for numerical calculations. 
\begin{enumerate}
\item A rubidium atom which has Loschmidt constant $n=2.687 \times 10^{25}\text{ \ } m^{-3}$, mass  $m=1.445\times 10^{-25}\text{ \ }kg$, and spin degeneracy $g=1$.

\item A sodium atom where $n=1.5 \times 10^{20}\text{ \ } m^{-3}$,  $m=3.819\times 10^{-26}\text{ \ }k g$, and $g=1$. 


\end{enumerate}

\begin{figure}[htb!]
\begin{minipage}[t]{0.5\textwidth}
        \centering
        \includegraphics[width=\textwidth]{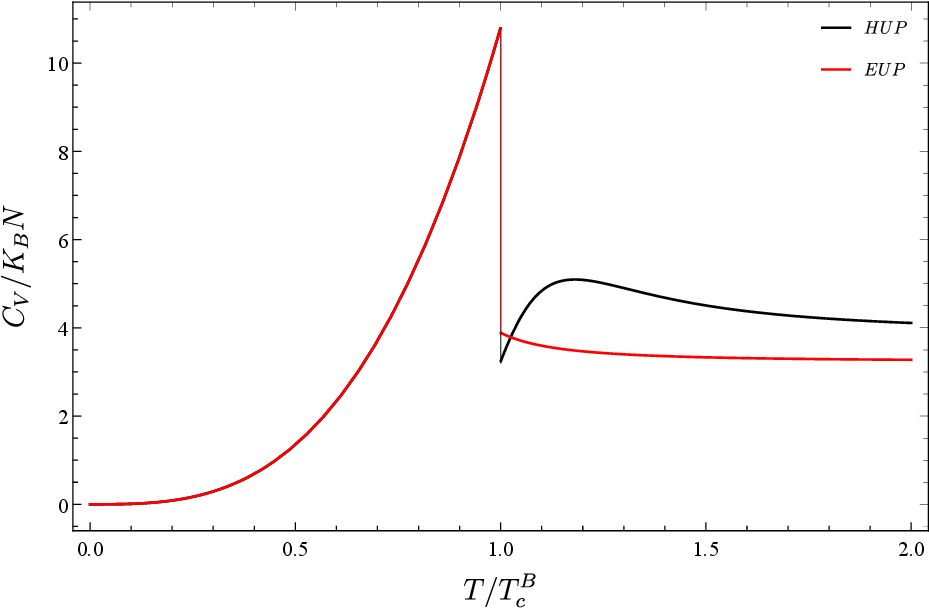}
        \label{fig:ha}
        \subcaption{Rubidium atom Bose gas}
\end{minipage}%
\begin{minipage}[t]{0.5\textwidth}
        \centering
        \includegraphics[width=\textwidth]{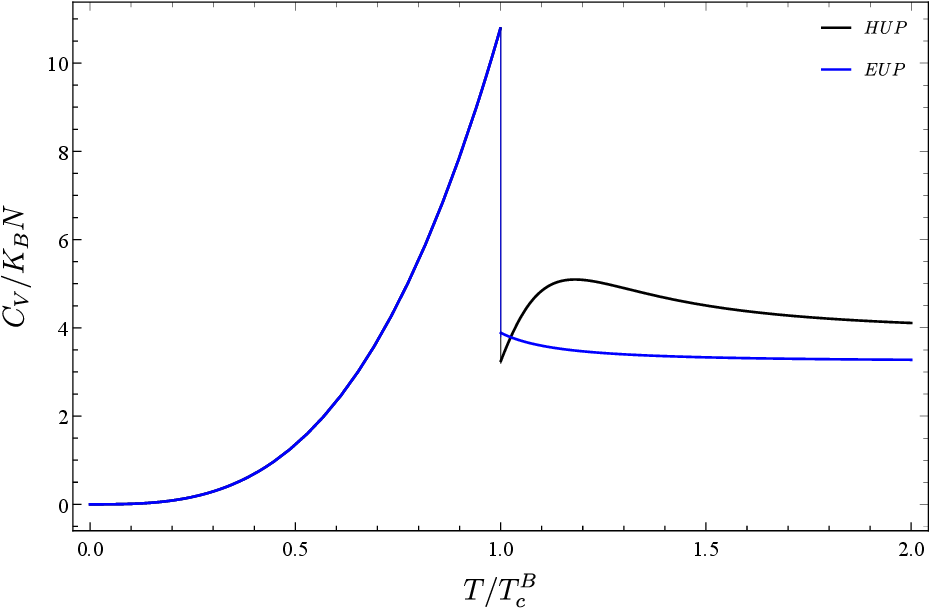}
       \label{fig:hb}
       \subcaption{Sodium atom Bose gas}
   \end{minipage}%
\caption{ The variations of $\frac{C_{V}}{K_{B}N}$  as function of $\frac{T}{T{c}^{B}}$.}
\label{fig3}
\end{figure}

It is worth noting that these dilute gases were originally used to verify BEC experimentally \cite{Anderson1995, Davis1995}. Using the numerical values, we plot the reduced specific heat function versus the temperature ratio in Fig. \ref{fig3}. 

\newpage
We then provide a comparison of the thermodynamic modifications in Table \ref{tab:1}. Here, we see that the EUP modification is very small to be detected at low temperatures. 
\begin{table}[htb!]
\centering
\begin{tabular}{l|l|l}
\hline\hline
& Rubidium & Sodium \\ \hline
$\frac{\Delta U}{N}$ (J/particle) & -7.55943$\times 10^{-37}$ & -1.11345$%
\times 10^{-36}$ \\ 
$\frac{\Delta C_{V}}{NK_{B}}$ (J/K particle) & -4.86241$\times 10^{-5}$ & 
-7.16201$\times 10^{-5}$ \\ 
\hline\hline
\end{tabular}
\caption{EUP modification estimation on the internal and specific heat functions for rubidium and sodium atomic gas systems trapped harmonically at $T=T_{c}^{B}$.}
\label{tab:1}
\end{table}

\section{Modified Ideal Fermi gas in Harmonic Traps}\label{sec3}

It is well known that, as a first approximation, nucleons in atomic structures and electrons in metallic substances can be modelled as an ideal Fermi gas. In this context, the scenario where the temperature approaches absolute zero is of particular importance. In this section, we study the main thermodynamic quantities of the ideal Fermi gas confined in a three-dimensional harmonic trap in the framework of EUP. We begin by defining the number of particles in a Fermi system
\begin{equation}
N=\sum_{k}\left\langle n_{k}\right\rangle =\sum_{k}\frac{1}{z^{-1}e^{\frac{%
E_{k}}{K_{B}T}}+1},  \label{f1}
\end{equation}%
where the summation is over all individual energy eigenstates. In contrast to the Bose scenario, in the Fermi case the parameter $z$ can take any value in the range $0\leq z\leq +\infty.$ Now we replace the discrete summation over $k$ by the corresponding integration, Eq. \eqref{f1} into 
\begin{equation}
N=\frac{g}{2}\left( \frac{K_{B}T}{\hbar \omega }\right)^{3}\int_{0}^{+\infty }\frac{y^{2}\left( 1-\frac{3\alpha K_{B}T}{m\omega ^{2}}y\right) dy}{z^{-1}e^{y}+1},  \label{f2}
\end{equation}%
where $g=2s+1$ is the degeneracy constant. In Eq. \eqref{f2} we see the appearance of the Fermi-Dirac function in the form of
\begin{equation}
f_{n}\left( z\right) =\frac{1}{\Gamma \left( n\right) }\int_{0}^{+\infty }\frac{y^{n-1}dy}{z^{-1}e^{y}+1}=\sum_{j=1}^{\infty }\left( -1\right) ^{j-1}\frac{z^{j}}{j^{n}}.
\end{equation}%
We then obtain the fermionic mean occupation number
in the presence of the EUP as follows: 
\begin{equation}
N=g\left( \frac{K_{B}T}{\hbar \omega }\right) ^{3}f_{3}\left( z\right)
\left( 1-\frac{9\alpha K_{B}T}{m\omega ^{2}}\frac{f_{4}\left( z\right) }{%
f_{3}\left( z\right) }\right) .  \label{fn}
\end{equation}%
Let us also emphasize that the $\alpha$ dependency is a natural consequence of modifying the Heisenberg algebra. It is also worth noting that the formula for the average occupation number can be contrasted with the special case where $\alpha =0$. In this scenario, we recover the precise result for the conventional ideal Fermi gas confined in harmonic traps. 

We now consider the thermodynamics of trapped Fermi gases, beginning with the calculation of the internal energy, 
\begin{equation}
U=\frac{g}{2\hbar ^{3}\omega ^{3}}\int_{0}^{+\infty }\frac{E^{3}\left( 1-%
\frac{3\alpha E}{m\omega ^{2}}\right) }{z^{-1}e^{\frac{E}{K_{B}T}}+1}dE.
\label{fe}
\end{equation}%
Using the Fermi-Dirac function, we see that Eq. \eqref{fe} becomes  
\begin{equation}
U=3g\left( \frac{K_{B}T}{\hbar \omega }\right) ^{3}K_{B}Tf_{4}\left(z\right) \left( 1-12\frac{\alpha K_{B}T}{m\omega^{2}}\frac{f_{5}\left(
z\right) }{f_{4}\left( z\right) }.\right)
\end{equation}%
We then employ Eq. \eqref{fn} to eliminate the term $\left( \frac{K_{B}T}{\hbar \omega }\right)^{3}$, so we get 
\begin{equation}
U=3NK_{B}T\frac{f_{4}\left( z\right) }{f_{3}\left( z\right) }\left( 1-\frac{12\alpha K_{B}T}{m\omega ^{2}}\frac{f_{5}\left( z\right) }{f_{4}\left(z\right) }+\frac{9\alpha K_{B}T}{m\omega ^{2}}\frac{f_{4}\left( z\right) }{f_{3}\left( z\right) }\right) .
\end{equation}%
It is clear that when $\alpha =0$, we have%
\begin{equation}
U=3NK_{B}T\frac{f_{4}\left( z\right) }{f_{3}\left( z\right) },
\end{equation}%
which is similar to the result obtained in dealing with the case of a Fermi gas in harmonic traps. On the other hand, in the classical limit where $z\rightarrow 0$, we can approximate $f_{n}\left( z\right) \sim z$, thus
\begin{equation}
U = 3NK_{B}T\left( 1-\frac{3\alpha K_{B}T}{m\omega ^{2}}\right) .
\end{equation}%
Here, the first term is the ordinary internal energy of the ideal Fermi gas in the harmonic trap and the second term is the correction due to the presence of the EUP.

Next, we derive another important thermodynamic quantity, the heat capacity at constant volume, which is directly related to the internal energy with the relation given in Eq. \eqref{cvv}. Straightforward calculation gives%
\small
\begin{eqnarray}
C_{V}=12gK_{B}\left( \frac{K_{B}T}{\hbar \omega }%
\right) ^{3}\left( f_{4}\left( z\right) -15\frac{\alpha K_{B}T}{m\omega
^{2}}f_{5}\left( z\right) \right) +3 g K_{B} T \left( \frac{K_{B}T}{\hbar \omega }%
\right) ^{3}\left( f_{3}\left( z\right) -12\frac{\alpha K_{B}T}{%
m\omega ^{2}}f_{4}\left( z\right) \right) \frac{1}{z}\frac{\partial z}{%
\partial T}.
\end{eqnarray}%
\normalsize
By utilizing Eq. \eqref{fn},  we eliminate the $\frac{1}{z}\frac{\partial z}{\partial T}$ term as
\begin{equation}
\frac{T}{z}\frac{\partial z}{\partial T}=-3\frac{f_{3}\left( z\right) }{%
f_{2}\left( z\right) }\left( 1+\frac{9\alpha K_{B}T}{m\omega ^{2}}\frac{%
f_{3}\left( z\right) }{f_{2}\left( z\right) }-\frac{12\alpha K_{B}T}{m\omega
^{2}}\frac{f_{4}\left( z\right) }{f_{3}\left( z\right) }\right) .  \label{fz}
\end{equation}%
Then, using Eqs. (\ref{fn}) and (\ref{fz}) we express the EUP-modified heat capacity at constant volume in the form of 
\begin{eqnarray}
\frac{C_{V}}{K_{B}N} &=&12\frac{f_{4}\left( z\right) }{f_{3}\left( z\right) }%
\left( 1+\frac{9\alpha K_{B}T}{m\omega ^{2}}\frac{f_{4}\left( z\right) }{%
f_{3}\left( z\right) }-15\frac{\alpha K_{B}T}{m\omega ^{2}}\frac{f_{5}\left(
z\right) }{f_{4}\left( z\right) }\right)  \notag \\
&&-9\frac{f_{3}\left( z\right) }{f_{2}\left( z\right) }\left( 1+\frac{%
9\alpha K_{B}T}{m\omega ^{2}}\frac{f_{3}\left( z\right) }{f_{2}\left(
z\right) }-15\frac{\alpha K_{B}T}{m\omega ^{2}}\frac{f_{4}\left( z\right) }{%
f_{3}\left( z\right) }\right) .  \label{fc}
\end{eqnarray}%
For $z\rightarrow 0$, Eq. (\ref{fc}) yields to the classical limit 
\begin{equation}
\frac{C_{V}}{K_{B}N}=3\left( 1-6\frac{\alpha K_{B}T}{m\omega ^{2}}\right), 
\label{heatcla}
\end{equation}%
which is analogous to the ideal Bose gas case, given in Eq. \eqref{hcla}. Here we observe that the curvature effects lead to a linear decrease of the specific heat function with temperature. 

In this subsection, we finally examine the equation of state. To this end, we express the pressure with \begin{equation}
P=\frac{K_{B}T}{V}\sum_{k}\ln \left( 1+ze^{\frac{-E_{k}}{K_{B}T}}\right) .
\end{equation}%
For a very large volume of phase space, the sum can be approximated by an integral in the pressure expression, and after integration, we obtain the following equation
\begin{equation}
\frac{PV}{NK_{B}T}=\frac{f_{4}\left( z\right) }{f_{3}\left( z\right) }\left(
1+\frac{9\alpha K_{B}T}{m\omega ^{2}}\frac{f_{4}\left( z\right) }{%
f_{3}\left( z\right) }-\frac{9\alpha K_{B}T}{m\omega ^{2}}\frac{f_{5}\left(
z\right) }{f_{4}\left( z\right) }\right) .  \label{fp}
\end{equation}%
Here we find that Eq. \eqref{fp} is completely equivalent to that for an ideal Bose gas as given in Eq. \eqref{bp}. Thus, in the classical limit $z\rightarrow 0,$ Eq. \eqref{fp} simplifies to the classical ideal gas law, $PV=NK_{B}T$.

\subsection{Degenerated Fermi gas}\label{subsec1}
We now aim to examine the opposite limiting scenario, where temperatures are low, and densities are high. In the extreme case of $T=0$, the fermionic mean occupation number $\left\langle n_{k}\right\rangle $ can be approximated by the step function $\theta \left( \mu -E\right) $, where
\begin{equation}
\theta \left( \mu -E\right) =\left\{ 
\begin{array}{c}
1\text{ \ if }\mu \geq E \\ 
0\text{ \ if }\mu <E%
\end{array}%
\right. .  \label{fs}
\end{equation}
At $T=0$, the chemical potential is equal to the Fermi energy, $E_{F}$, of the system. In this case, using Eq. \eqref{fs} we can directly calculate the number of particles
\begin{eqnarray}
N &=&\frac{g}{2\hbar ^{3}\omega ^{3}}\int_{0}^{E_{F}}E^{2}\left( 1-\frac{3\alpha E}{m\omega ^{2}}\right) dE=\frac{g}{6\hbar ^{3}\omega ^{3}}\left( E_{F}^{3}-\frac{9\alpha E_{F}^{4}}{4m\omega ^{2}}\right) .  \label{ff}
\end{eqnarray}%
Therefore, we can express the EUP-modified Fermi energy to first order in $\alpha $ as: 
\begin{equation}
E_{F}=E_{F0}\left( 1+\frac{3\alpha }{4m\omega ^{2}}E_{F0}^{3}\right) ,
\label{EF}
\end{equation}%
where 
\begin{equation}
E_{F0}=\left( \frac{6\hbar ^{3}\omega ^{3}N}{g}\right) ^{1/3}.  \label{EEF}
\end{equation}%
is the conventional Fermi energy. 

At this point, we employ the step function to derive the internal energy. In this case, the EUP-modified ground state energy reads:
\begin{equation}
U_{0}=\frac{g}{2\hbar ^{3}\omega ^{3}}\int_{0}^{E_{F}}E^{3}\left( 1-\frac{%
3\alpha E}{m\omega ^{2}}\right) dE=\frac{gE_{F}^{4}}{8\hbar ^{3}\omega ^{3}}%
\left( 1-\frac{12\alpha E_{F}}{5m\omega ^{2}}\right) .
\end{equation}%
Thus, using Eq. \eqref{ff} we express the EUP-modified ground state energy per particle as follows
\begin{equation}
\frac{U_{0}}{N}=\frac{3}{4}E_{F0}\left( 1-\frac{3\alpha }{20m\omega ^{2}}%
E_{F0}+\frac{3\alpha }{4m\omega ^{2}}E_{F0}^{3}\right) .
\end{equation}%
We now want to calculate the correction to this limit at low temperatures. For small but non-zero temperatures, the Fermi integration can be expressed as a rapidly converging series using the Sommerfeld lemma \cite{Huang, Pathria}
\begin{equation}
f_{n}\left( z\right) =\frac{\left( \ln z\right) ^{n}}{n!}\left[ 1+n\left(
n-1\right) \frac{\pi ^{2}}{6}\frac{1}{\left( \ln z\right) ^{2}}+n\left(
n-1\right) \left( n-2\right) \left( n-3\right) \frac{7\pi ^{2}}{360}\frac{1}{%
\left( \ln z\right) ^{4}}+...\right] .  \label{fzz}
\end{equation}%
After we substitute Eq. \eqref{fzz} into Eq. \eqref{fn}, we get
\begin{equation}
N=\frac{g\mu ^{3}}{6\hbar ^{3}\omega ^{3}}\left[ 1+\pi ^{2}\left( \frac{K_{B}T}{\mu }\right) ^{2}-\frac{9\alpha \mu }{4m\omega ^{2}}-\frac{9\pi^{2}\alpha \mu }{2m\omega ^{2}}\left( \frac{K_{B}T}{\mu }\right) ^{2}\right] ,
\end{equation}%
or alternatively 
\begin{equation}
\mu =E_{F0}\left[ 1+\pi ^{2}\left( \frac{K_{B}T}{\mu }\right) ^{2}-\frac{%
9\alpha \mu }{4m\omega ^{2}}-\frac{9\pi ^{2}\alpha \mu }{2m\omega ^{2}}%
\left( \frac{K_{B}T}{\mu }\right) ^{2}+...\right] ^{-1/3}.
\end{equation}
In the first approximation, the chemical potential is 
\begin{equation}
\mu =E_{F0}\left[] 1+\frac{3\alpha E_{F0}}{4m\omega ^{2}}-\frac{\pi ^{2}}{3}%
\left( \frac{T}{T_{F0}}\right) ^{2}+\frac{3\pi ^{2}\alpha E_{F0}}{2m\omega
^{2}}\left( \frac{T}{T_{F0}}\right) ^{2}\right] ,
\end{equation}%
where $T_{F0}=\frac{E_{F0}}{K_{B}}.$ We then rewrite the internal energy as
\begin{equation}
\frac{U}{N}=\frac{3}{4}E_{F0}\left[ 1-\frac{12\alpha E_{F0}}{5m\omega ^{2}}%
+2\pi ^{2}\left( \frac{T}{T_{F0}}\right) ^{2}-\frac{6\pi ^{2}\alpha E_{F0}}{%
m\omega ^{2}}\left( \frac{T}{T_{F0}}\right) ^{2}\right] .
\end{equation}
Finally, with the help of Eq. (\ref{cvv}), we obtain the heat capacity in the following form 
\begin{equation}
\frac{C_{V}}{K_{B}N}=3\pi ^{2}\left( 1-\frac{3\alpha E_{F0}}{m\omega ^{2}}%
\right) \left( \frac{T}{T_{F0}}\right) .  \label{hff}
\end{equation}
In particular,  Eq. \eqref{hff} indicates that for $\frac{T}{T_{F0}}<<1$ the heat capacity increases in a linear fashion with temperature, reaching zero when the temperature attains absolute zero. Furthermore, EUP introduces a negative adjustment to this behaviour that results in reducing the thermal capacity of the system. To illustrate this, let us consider a copper-electron gas in a harmonic potential. Copper has a mass density of $8.9\times 10^{3}kg/m^{3}$, an atomic weight of $63$, and an electron has a mass of $m_{e}=9.1\times 10^{-31} kg$. Assuming that  each copper atom provides one free electron, the electron number density can be calculated. In this case, the number of electrons per unit volume is given by $n=\frac{N}{V}=8.5\times 10^{28}m^{-3}$. The Fermi energy of the copper-electron gas in a harmonic potential is $ E_{F0}=1.9828 \times 10^{-9}$ $\text{J}$ \cite{HLi2018}, and the amendments to the Fermi energy due to EUP is $\Delta E_{F}=9.7096\times 10^{-25}$ $\text{J}$ which is very small compared to $E_{F0}$. It is important to note that the EUP has a significantly smaller impact on the copper-electron gas in a harmonic potential compared to the GUP. The magnitude of the EUP's effect, which is order of $10^{-25}$, is negligible compared to $1$. Consequently, when studying a copper-electron gas in a harmonic potential at low temperatures, the effects of the EUP can be safely disregarded due to their minimal influence on the system's behavior.

\section{Conclusion} \label{conc}

In this manuscript, we investigate the curvature effects on ideal Bose and Fermi gases in the framework of EUP. We find that in the ideal Bose gas system, the curvature effect sets an upper bound value on the temperature and it alters the conventional condensation temperature by a positive correction term. We then observe that the curvature effect does not significantly change the ratio of the ground and thermal state particle numbers to the total particle number when the temperature is relatively very small than the critical temperature. We then derive the internal energy and show that the curvature influence has a reducing impact on the internal energy. Next, we obtain the specific heat and equation of the state. Below the critical temperature, both functions present a decrease as the curvature effects increase. 

Next, we study the ideal Fermi gas in the context of the EUP. First, we show how the average occupation number depends on the EUP deformation parameter. We then derive the EUP-modified internal energy and show that in the $z\rightarrow 0$ limit, the curvature correction quadratically suppresses the linear increase in internal energy due to tempering by temperature. Then, we employ the internal energy to obtain the EUP-modified heat capacity at constant volume. We find that in the classical limit the curvature effects cause a decrease in specific heat linear in temperature. After that we derive the specific heat function and we observe that in the classical limit the EUP modification causes the specific heat function to decrease linearly with temperature. We then show that the EUP-modified equation of state function in ideal Fermi and Bose gases are the same. 

Finally, we have studied the effect of the EUP formalism on the degenerate Fermi gas. First, we obtain the EUP-modified Fermi energy up to the first order correction and express it in terms of the conventional one. From this we derive the EUP-modified ground state energy per particle and then express it in series expansion in terms of non-zero small temperatures. Finally, we construct the EUP-modified specific heat function which reveals the linear decreasing effect of temperature. As an example, we discuss our results by considering a copper electron gas in a harmonic potential at low temperatures. We find that the effect of the EUP formalism can be safely neglected, since the modification on the behaviour of the system is very small.

\section*{Acknowledgments}

This work is supported by the Ministry of Higher Education and Scientific Research, Algeria under the code: B00L02UN040120230003. { B. C. L. is grateful to Excellence Project PřF UHK 2211/2023-2024 for the financial support.}

\end{document}